\documentclass[twocolumn,superscriptaddress,showpacs,amsfonts]{revtex4-1}
\pdfoutput=1

\usepackage{epsfig}
\usepackage{graphicx}

\newcommand{\bea}{\begin{eqnarray}}
\newcommand{\eea}{\end{eqnarray}}
\newcommand{\bt}{\textbf}
\newcommand{\mb}{\mathbf}
\newcommand{\phd}{\phantom{\dag}}
\newcommand{\ph}{\phantom{.}}
\newcommand{\up}{^{\phd}}
\newcommand{\noi}{\noindent}

\begin{document}


\title{Magnetic manipulation of topological states in p-wave superconductors}

\author{Maria~Teresa Mercaldo}
\affiliation{Dipartimento di Fisica ``E. R. Caianiello'',
Universit\`a di Salerno, I-84084 Fisciano (Salerno), Italy}
%
\author{Mario Cuoco}
\affiliation{CNR-SPIN, I-84084 Fisciano (Salerno), Italy}
\affiliation{Dipartimento di Fisica ``E. R. Caianiello'',
Universit\`a di Salerno, I-84084 Fisciano (Salerno), Italy}
\author{Panagiotis Kotetes}
\affiliation{Center for Quantum Devices, Niels Bohr Institute, University of Copenhagen, 2100 Copenhagen, Denmark}

\begin{abstract}
Substantial experimental investigation has provided evidence for spin-triplet pairing in diverse classes of materials and in a variety of artificial heterostructures. A fundamental challenge in actual experiments is how to manipulate the topological behavior of $p$-wave superconductors (PSCs) that could open perspectives for applications. Such a control knob is naturally provided by the spin-triplet character of the PSC order parameter, described by the spin $\mb{d}$-vector. Therefore, in this work we investigate the magnetic field response of one-dimensional (1d) PSCs and demonstrate that the structure of the Cooper pair spin-configuration is crucial to set topological phases with an enhanced number of Majorana fermions per edge, ${\cal N}$, ranging from ${\cal N}=0$ to $4$. The topological phase diagram, consisting of phases with Majorana modes at the edge, becomes significantly modified when one tunes the strength of the applied field and allows for long range hopping amplitudes in the 1d PSC. We find transitions between phases with different number of Majorana fermions per edge that can be both induced by a variation of the hopping strength and a spin rotation of $\mb{d}$. Hence, the interplay of the applied magnetic field and the internal spin degree of freedom of the PSC opens a new promising route for engineering topological phases with large number of Majorana modes. 
\end{abstract}
\maketitle
\noindent 



\section{Introduction}
\label{sec1}

A Majorana fermion (MF) is an exotic quasiparticle that constitutes its own charge-conjugate partner (i.e. creation and annihilation operators are identical)~\cite{Majorana1937}. This property implies that every spinless fermion can be decomposed into two MFs, locally bound together. However, if the two MFs becomes sufficiently spatially separated, then one can employ the two states defined by the respective spinless fermion as a topological qubit. The experimental achievement and control of the above pro\-per\-ties are timely challenges in the area of condensed matter from both a fundamental point of view~\cite{Wilczek2009} and also for the tantalizing perspectives of decoherence free quantum computation~\cite{Kitaev2001,Kitaev2003,Nayak2008}. Recently, there has been a tremendous effort in designing and realizing materials platforms where topological superconductivity and MFs can be successfully obtained. This is demonstrated by the proposal of a large variety of systems based on he\-te\-ro\-structures made of topological insulators or semiconductors interfaced with $s$-wave superconductors (SCs)
\cite{FuKane2008,AliceaReview,BeenakkerReview,FlensbergReview,KotetesClassi} as well as by the subsequent experimental evidence of MFs in hybrid superconducting devices \cite{Mourik,Deng,Furdyna,Heiblum,Finck,Churchill,Franke,Pawlak,Albrecht,Sole}.

Apart from artificial systems, natural candidates for MFs are also intrinsic $p$-wave superconductors where electrons form Cooper pairs in symmetric spin-triplet and orbital-antisymmetric configurations. The number of known triplet superconductors has been growing steadily, and now includes UPt$_3$~\cite{Stewart1984},
ferromagnetic superconductors such as UGe$_2$, URhGe, UIr and UCoGe~\cite{Saxena2000,Aoki2001,Akazawa2004},
quasi one-dimensional organic system (TMTSF)$_2$X (X=ClO$_4$ and PF$_6$)~\cite{Lebed1986,Lee1997} and transition metal lithium oxide~\cite{Greenblatt1984}. 
In particular, hallmarks for spin-triplet superconductivity have emerged for Sr$_2$RuO$_4$~\cite{Maeno2012,MaenoRMP}, for which the existence of high-quality single crystals allowed the study of heterostructures based on spin-triplet superconductivity~\cite{Anwar2014,Gentile2013}.

While the so far conducted studies of PSCs have led to remarkable progress and insight regarding topological systems at large, some fundamental aspects remain not fully established. For instance, one of the main issue when dealing with PSC points to the relation between the spin-structure of the triplet order parameter (OP) and the resulting topological phases.

\begin{figure}[t]
\begin{center}
\includegraphics[width=1.0\columnwidth]{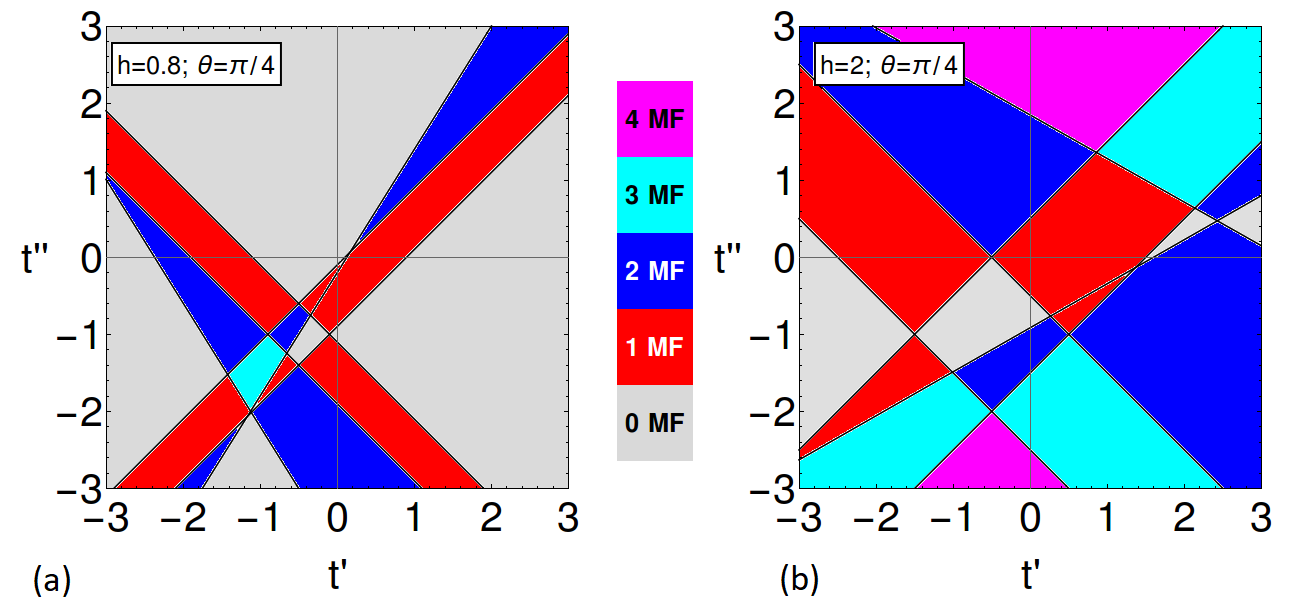}
\end{center}
\caption{Topological phase diagram in the $[t',t'']$ plane for a representative set of parameters: $\theta=\pi/4$, $V=-2$, $\mu=1.0$, $\Delta_{\uparrow\uparrow}=0.09$, $\Delta_{\downarrow\downarrow}=0.9$ and two magnetic field amplitudes (a) $h=0.8$ and (b) $h=2.0$. All energies are in unit of $t$. Colors refer to regions with a different number of MFs per edge.  A sufficiently large value of the applied field is needed to get both topological phases with 4 MFs per edge as well as to have topological states almost independently on the choice of $[t',t'']$ plane in a wide range of amplitudes. Topological transitions from 1 to 3 MFs per edge are mostly accessible in high field.}
\label{fig:1}
\end{figure}

In this work, we move in this framework with a special focus on the possibility of achieving topological phases with a large number of MFs per edge. We show that the Cooper pair spin-configuration of a 1d PSC with an easy spin-plane, chiral symmetry \cite{Dumitrescu2014,Tewari and Sau, ChiralTanaka,SatoChiral,NOZeemanPK,Mercaldo2016} and long-range hoppings can have a fundamental role to set topological phases with an enhanced number of Majorana fermions per edge (e.g. ranging from ${\cal N}=0$ to $4$).
We determine the topological phase diagram, consisting of phases with MFs at the edge, in the presence of long range hopping amplitudes in the 1d PSC and an applied magnetic field that preserves the chiral symmetry. We find transitions between phases with different number of Majorana fermions per edge that can be both induced by a va\-ria\-tion of the hopping strength and a modification of the $\bm{d}$-vector spin structure. Hence, magnetic field and internal spin degree of freedom of the PSC define relevant tuning parameters to engineer topological phases with large number of MFs.

The remainder of the paper is organized as follows. We present in Sect.~\ref{sec2} the model and the methodology for describing the 1d PSC. Sect.~\ref{sec3} is devoted to the presentation of the results as related to the topological phase diagrams. Finally, in the Sect.~\ref{sec4} we provide the concluding remarks.

\section{Model and Methodology}
\label{sec2}

We model the 1d-PSC by a Bogoliubov-de Gennes lattice Hamiltonian in $k$-space as,
\bea
\widehat{{\cal H}}_k=\varepsilon_k\tau_z-\bm{h}\cdot\bm{\sigma}+\tau_+\ph\bm{d}_k\cdot\bm{\sigma}+\tau_-\ph\bm{d}_k^*\cdot\bm{\sigma}\,,\quad\label{eq:Hamiltonian}
\eea

\noi where ${\cal H}=\frac{1}{2}\sum_k\Psi_k^{\dag}\widehat{\cal H}_k\Psi_k$ in the Nambu basis $\Psi_k^{\dag}=(\psi_{k\uparrow}^{\dag}\,,\psi_{k\downarrow}^{\dag}\,,\psi_{-k\downarrow}\up\,,-\psi_{-k\uparrow}\up)$. The matrices $\bm{\sigma}$ and $\bm{\tau}$ indicate spin 1/2 Pauli matrices in the spin and particle-hole sectors, respectively. 
We assume the electron dispersion $\varepsilon_k=-2t\cos(ka)-2t'\cos(2ka)-2t''\cos(3ka)-\mu$ and $t=1$, with $t^{\nu}$ denoting hopping to the $\nu$-th nearest neighbor with lattice constant $a=1$ and $\mu$ is the che\-mi\-cal potential. In addition, we introduce the Zeeman field $\bm{h}$ and the odd-parity OP $\bm{d}_k=2\bm{d}\sin k$, with $\bm{d}$ the complex vector defining the spin-orientation of the OP. It is also convenient to introduce a matrix OP in spin-space $\{\uparrow,\downarrow\}$, $\widehat{\Delta}=\bm{d}\cdot\bm{\sigma}\sigma_y$: $\Delta_{\uparrow\uparrow,\downarrow\downarrow}=d_y\pm id_x$ and $\Delta_{\uparrow\downarrow}=-id_z$. The $\bm{d}$-vector components are then related to the pairing correlations for the spin-triplet configurations having zero spin projection along the corresponding symmetry axis. Moreover, once we know the orientation of the $\bm{d}$-vector we can immediately deduce that Cooper pairs having equal spin configurations are in the plane perpendicular to it.

For the present analysis we consider a magnetic field lying in the $yz$ plane with amplitude $h$ and orientation $\theta$, i.e. $h_y=h \sin[\theta]$, $h_y=h \cos[\theta]$), while the PSC has an easy $xy$ spin-plane for the $\bm{d}$-vector, and in particular our previous self-consistent analysis~\cite{Mercaldo2016} has shown that $\bm{d}=d(i\cos\alpha,\sin\alpha,0)$. The spin structure of the pai\-ring is due to an effective separable four-fermion inte\-rac\-tion in the PSC channel with potentials $V_x=V_y\equiv V$ for the spin- $xy$ plane.
For such a physical configuration, although in the presence of a source of the usual time reversal symmetry brea\-king, at any given orientation of the field in the $yz$ plane the Hamiltonian resides in the BDI symmetry class exhi\-bi\-ting chiral, time-reversal and charge-conjugation symmetries \cite{Altland,KitaevClassi,Ryu} with corre\-spon\-ding ope\-ra\-tors: $\Pi=\tau_x\sigma_x$, $\Theta=\tau_z\sigma_z{\cal K}$ and $\Xi=\tau_y\sigma_y{\cal K}$. 
Here, ${\cal K}$ stands for complex conjugation. Note that the emerging time-reversal symmetry does not lead to a Kramers degeneracy, since it satisfies $\Theta^2=I$, with $I$ the unit operator. 
\begin{figure}[t]
\begin{center}
\includegraphics[width=1.0\columnwidth]{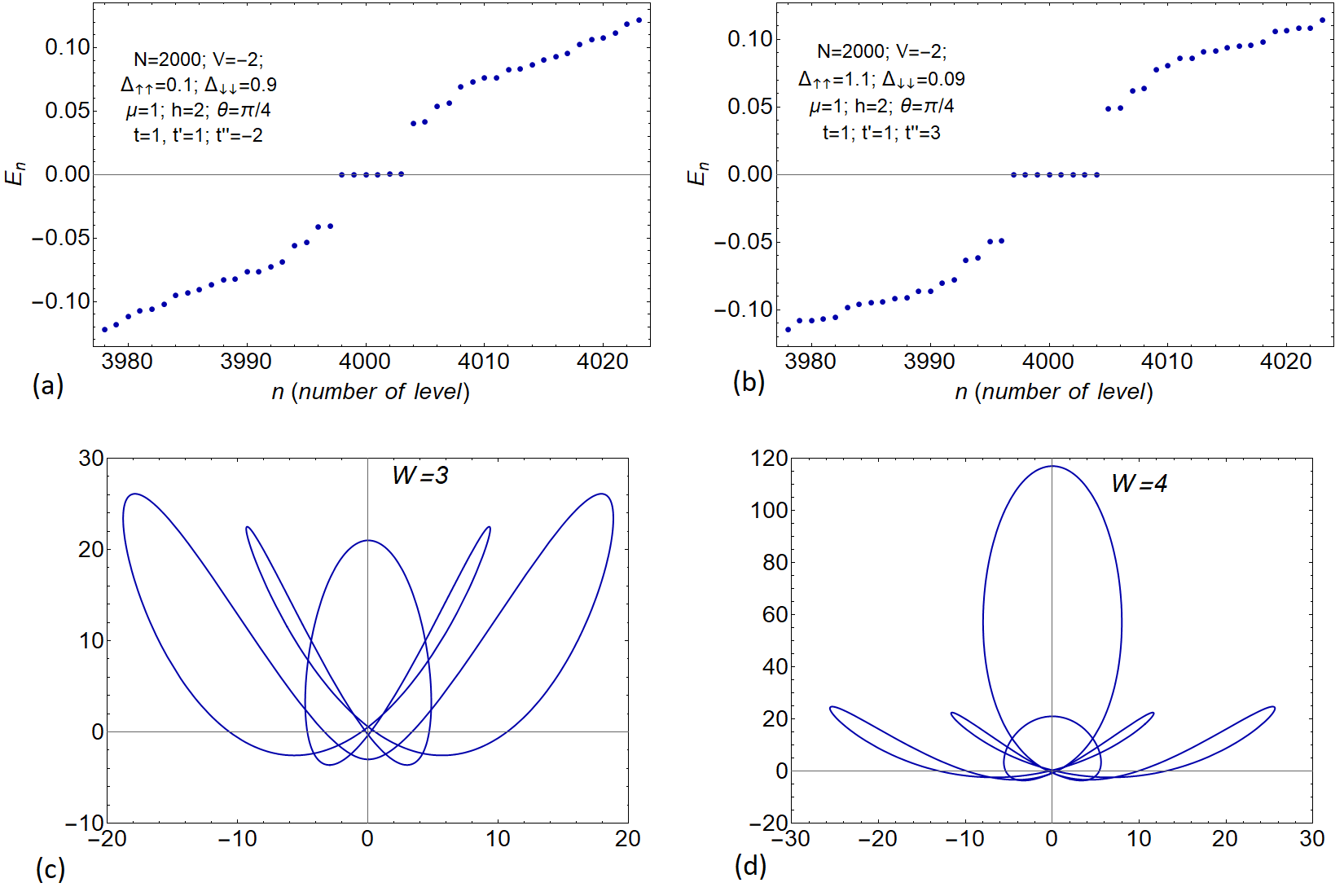}
\end{center}
\caption{(a)-(b) near-zero low energy spectrum for an open chain of $N=2000$ sites for a set of parameters that yield a topological phase with 3 and 4 MFs per edge, respectively. (c)-(d) parametric plots of the ${\rm Re}[\det(A_k)]$ and ${\rm Im}[\det(A_k)]$ as the momentum $k$ is varied in the Brillouin zone (i.e. $k\epsilon [-\pi,\pi]$). The trajectory in the complex plane indicates that the curve winds around the origin three (c) and four (d) times thus yielding a winding number $W$=3 and 4, respectively.}
\label{fig:2}
\end{figure}
The coexistence of the above set of symmetries is very important because it sets the topological class of the system and it allows to determine an integer $\mathbb Z$ invariant which counts the number of topologically protected MFs at the end of the 1d-PSC. Since the chiral symmetry operator anti-commutates with $\widehat{{\cal H}}_k$, by employing a unitary transformation $U$ rotating the basis in the eigenbasis of $\Pi$, the Hamiltonian can be put in the off-diagonal form as
\bea
\widehat{{\cal H}}^{'}_k= U \widehat{{\cal H}} U^{\dagger}=\left(\begin{array}{cc}
0 & A_{k}\\
A_{-k}^{\dagger} & 0
\end{array}\right).  \label{eq:Hamiltonian1}
\eea
with antidiagonal blocks given by the matrices $A_{k}$. Hence, its determinant
$\det{A_{k}}$ can be put in a complex polar form $z_k= |\det{A_{k}}| \exp[i \theta_k]$ and, as long as the eigenvalues of ${A}_{k}$ are non-zero, it can be used to obtain the winding number $W$ by evaluating its trajectory in the complex plane as
\bea
W= \frac{1}{2 \pi i} \int_{-\pi}^{\pi} \frac{d z_k}{z_k} \,. \label{eq:Hamiltonian2}
\eea
We observe that the number of singularities in the phase of the determinant is a topological invariant~\cite{Tewari and Sau} because it is not
related to any symmetry breaking and it cannot change without the amplitude going to zero, thus implying a gap closing 
and a topological phase transition. The integer $W$ counts the number of MFs at the edges of 1d-PSC. Hereafter, we compute $W$ in the parameters space and we combine such analysis with that on an open chain of size $N=2000$~\cite{Cuoco1,Cuoco2} in order to investigate topological phases and topological transitions.   

\section{Results}
\label{sec3}

Since at zero applied field one can freely choose the orientation of the $\bm{d}$-vector, the Hamiltonian can be decomposed into two spin blocks, each of which contributes with 1MF per edge if $|t+t''|>|t'+\mu/2|$. The application of an external field (or effectively the proximity to a ferromagnetic system) leads to different behaviors depending on its orientation with respect to the $\bm{d}$-vector, including the possibility of a breakdown of the bulk-boundary correspondence due to a reconstruction of the bulk $\bm{d}$-vector arising from boundary effects~\cite{Mercaldo2016}. Moreover, while a field parallel to the $\bm{d}$-vector makes the PSC topologically trivial, for a perpendicular orientation a topological regime can be obtained with $1$ or $2$ MFs per edge. As mentioned in Sect. \ref{sec1}, the case with a magnetic field lying in a plane perpendicular to the $\bm{d}$-vector places the system in the class BDI and, in principle, it allows us to realize a topological phase with an arbitrary number of MFs per edge.  Starting from these observation, one anticipates a substantial reconstruction of the topological phase diagram in the presence of longer range neighbor tight-binding dispersions. In this context, our aim is to demonstrate that the spin-structure of the $\bm{d}$-vector in the chiral symmetric regime opens the path in obtaining new phases with an enhanced number of MFs per edge.

In Fig. \ref{fig:1}(a),(b) we report on the topological phase diagrams for representative sets of parameters for both the magnetic field and $\{\Delta_{\uparrow\uparrow},\Delta_{\downarrow\downarrow}\}$ OPs, with the latter corresponding to a two-components $\bm{d}$-vector 
with almost equal amplitude in 
the $xy$ spin-plane.
The construction of the topological phase diagram has been performed by directly computing the winding number $W$ through the trajectories in the complex plane of the $Det(A_k)$ as well as, for specific points in the phase space, by explicitly analyzing the multiple MFs in a 1d lattice with open boundary (see Figs. \ref{fig:2} (a)-(d)). The first observation arising from a direct comparison of the two phase diagrams at $h=0.8$ and $2.0$ is that topological phases with a number $\cal{N}$ of MFs larger than 3 requires a magnetic field that overcomes a critical threshold. Indeed, a large portion of the phase diagram in Fig. \ref{fig:1}(a) is topologically trivial (gray region) while configurations with ${\cal{N}}=3$ MFs are possible only by suitably tuning the $t'$ and $t''$ hopping amplitudes. The increase of the magnetic field not only leads to a dramatic growth of the topological regions with ${\cal{N}}=1,2,3$ MFs but it also gives access to new configurations with ${\cal{N}}=4$ MFs. Transitions between phases with different number of MFs can then be induced by both tuning the strength of the applied field and that of the $t'$ and $t''$ hopping amplitudes. We point out that it is not needed to have both $t'$ and $t''$ non-vanishing in order to get topological phases with ${\cal{N}}=3,4$ MFs. Nevertheless, it turns out that they are only accessible when $t''\neq0$ (Fig. \ref{fig:1}(b)), while, on the contrary, for $t''=0$ the winding number is limited to be less than 3.     

\begin{figure}[t]
\begin{center}
\includegraphics[width=1.0\columnwidth]{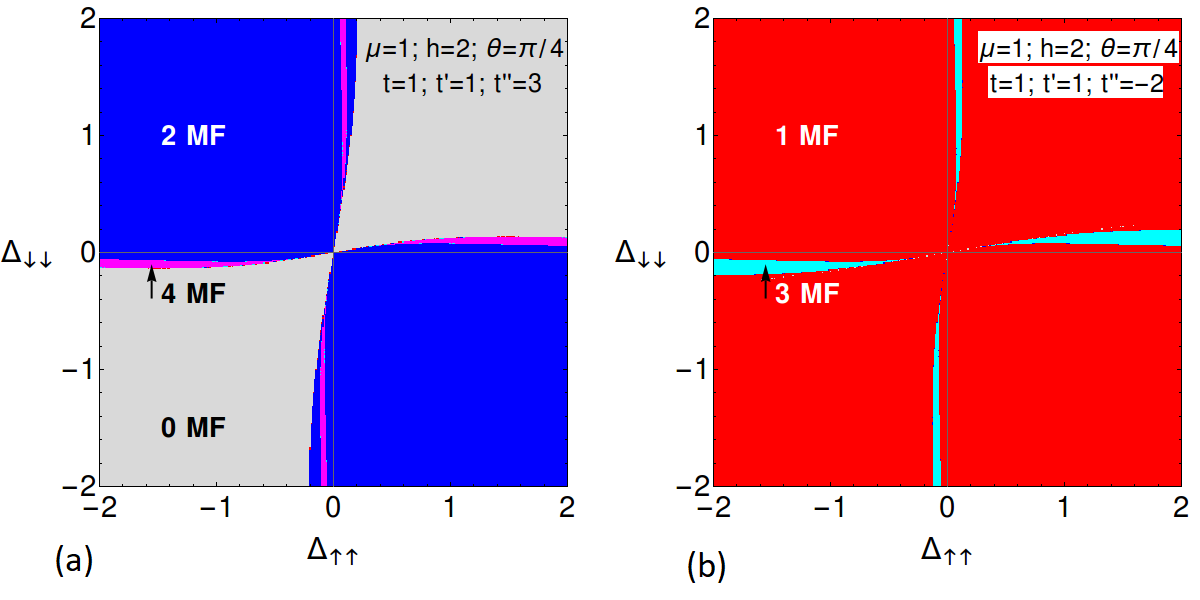}
\end{center}
\caption{Topological phase diagram in the $[\Delta_{\uparrow\uparrow},\Delta_{\downarrow\downarrow}]$ plane for $t''=3$ (a) and (b) $t''=-2$ for a given set of parameters: $t'=1$, $h=2$, $\theta=\pi/4$, $V=-2$, and $\mu=1.0$. All energies are in unit of $t$. Colors refer to regions with a different number of MFs per edge. We observe topological phases with 3 or 4 MFs that are accessible when the $\bm{d}$-vector  has almost equal 
amplitude for $x$ and $y$ components }
\label{fig:3}
\end{figure}

Finally, we address the role of the $\bm{d}$-vector by exploring a variation of the $\{\Delta_{\uparrow\uparrow},\Delta_{\downarrow\downarrow}\}$ OPs. To highlight the impact of the $\bm{d}$-vector rotation in the $xy$ plane we selected two representative configurations where all the electronic parameters are fixed apart from the amplitude of $t''$ (Fig. \ref{fig:3}). The resulting diagram provide interesting and general indications on the way the $\bm{d}$-vector acts in designing the topological phases. Indeed, we find that states with ${\cal{N}}=3,4$ MFs can be obtained only when $\Delta_{\uparrow\uparrow} \gg (\ll) \Delta_{\downarrow\downarrow}$ that corresponds
to configurations with the $\bm{d}$-vector having almost equal 
amplitude for $x$ and $y$ components.
Remarkably, such regions with enhanced MFs require to have both components of the $\bm{d}$-vector to be non-vanishing. Hence, we learn that in a field-temperature phase diagram they can potentially occur close by the boundary where one of the two spin order parameters goes to zero. 

When considering configurations with ${\cal{N}}=1,2$ MFs, as expected, those with an even number of MFs are sensitive to the relative sign of the OPs exhibiting a non-trivial behavior when a $\pi$ phase shift occurs between $\Delta_{\uparrow\uparrow}$ and $\Delta_{\downarrow\downarrow}$. On the other hand, the phases with an odd number MFs is not spin-phase dependent. 
An inequivalent behavior is also observed for the cases with ${\cal{N}}=3$ and $4$ MFs through different types of accessible topological phase transitions when interchanging $\Delta_{\uparrow\uparrow}$ with $\Delta_{\downarrow\downarrow}$. Our results show that phases with ${\cal{N}}=2$ and ${\cal{N}}=4$ have a fundamentally different topological behavior when considering their robustness and the nature of the allowed topological transitions upon a variation of the OPs amplitudes.

\section{Conclusions}
\label{sec4}

In conclusion we have investigated the relation between the spin-structure in 1d PSC and the occurrence of multiple MFs in the presence of an applied magnetic field and considering an electron connectivity with long-range hopping amplitudes. We demonstrated that for specific orientations of the magnetic field with respect to the $\bm{d}$-vector, chiral symmetry can be restored and the system can exhibit multiple chiral symmetry protected MFs. We singled out the main conditions which are needed to achieve such topological phases: the $\bm{d}$-vector should have two components with almost equal amplitude in 
the $xy$ spin-plane and thus it is generally non-collinear with respect 
to the orientation of the applied ﬁeld.

\end{document}